\documentclass[journal,twocolumn,letterpaper]{IEEEJERM}
\ifCLASSINFOpdf
\else
\fi

\usepackage{times,amsmath,epsfig}
\usepackage{fancyhdr}
\usepackage{amsmath}
\usepackage{amsfonts}
\usepackage{amssymb}
\usepackage[utf8]{inputenc}
\usepackage{array}
\usepackage{graphicx}
\usepackage{subcaption} 

\usepackage{bm}
\usepackage{breqn}
\usepackage{xcolor}
\usepackage{soul}
\usepackage{amssymb}
\usepackage[hyphens]{url}  
\usepackage{cite}
\usepackage{caption} 
\usepackage{multirow}
\usepackage{booktabs}
\usepackage{colortbl}
\usepackage{tabularx}
\usepackage{wrapfig}
\usepackage{float}

\usepackage[hidelinks]{hyperref}
\usepackage[most]{tcolorbox}  
\usepackage{orcidlink}
\usepackage[acronyms,nonumberlist,nopostdot,nomain,nogroupskip]{glossaries}
\makeglossaries
\glsdisablehyper
\newacronym{FPGA}{FPGA}{field-programmable gate array}
\newacronym{CNN}{CNN}{convolutional neural network}
\newacronym{NN}{NN}{neural network}
\newacronym{AI}{AI}{artificial intelligence}
\newacronym{ML}{ML}{machine learning}
\newacronym{DL}{DL}{deep learning}
\newacronym{SoC}{SoC}{system-on-a-chip}
\newacronym{GPU}{GPU}{graphics processing unit}
\newacronym{VPU}{VPU}{vision processing unit}
\newacronym{MS}{MS}{multispectral}
\newacronym{HS}{HS}{hyperspectral}
\newacronym{BoA}{BoA}{bottom-of-atmosphere}
\newacronym{SCL}{SCL}{scene classification layer} 
\newacronym{ESA}{ESA}{European Space Agency} 
\newacronym{CPU}{CPU}{central processing unit} 
\newacronym{EO}{EO}{Earth observation}
\newacronym{FDIR}{FDIR}{fault detection, isolation, and recovery}
\newacronym{COTS}{COTS}{commercial off-the-shelf}
\newacronym{LEO}{LEO}{Low Earth orbit}
\newacronym{HW}{HW}{hardware}
\newacronym{DPU}{DPU}{deep learning processor unit}
\newacronym{FP}{FP}{false positive}
\newacronym{FN}{FN}{false negative}
\newacronym{VAI}{VAI}{Vitis AI}
\newacronym{float32}{float32}{32-bit floating-point}
\newacronym{int8}{int8}{8-bit integer}
\newacronym{QAT}{QAT}{quantization-aware training}
\newacronym{FCNN}{FCNN}{fully convolutional NN}
\newacronym{FLOPs}{FLOPs}{floating-point operations per second}
\newacronym{PR}{PR}{pruning ratio}
\newacronym{IP core}{IP core}{intellectual property core}
\newacronym{MAC}{MAC}{multiply-accumulate}
\newacronym{PTQ}{PTQ}{post-training quantization}
\newacronym{IR}{IR}{Intermediate Rapresentation}
\newacronym{muC}{\textmu C}{microcontroller}
\newacronym{FPS}{FPS}{frame-per-second}
\newacronym{EC}{EC}{edge-computing}

\usepackage{comment}
\usepackage{newfloat}
\usepackage{listings}
\usepackage{xcolor}  

\newcommand{\highlightedurl}[1]{%
  \begingroup
  \hypersetup{pdfborder={0 0 1}}  
  \url{#1}
  \endgroup
}

\begin{document}

%
\title{Efficient FPGA-accelerated Convolutional Neural Networks for Cloud Detection on CubeSats}

%
\author{Angela~Cratere,~\IEEEmembership{Student~Member,~IEEE,}
        M.~Salim~Farissi,
        Andrea~Carbone,
        Marcello~Asciolla,~\IEEEmembership{Student~Member,~IEEE,}
        Maria~Rizzi, 
        Francesco~Dell'Olio,~\IEEEmembership{Senior~Member,~IEEE}, 
        Augusto~Nascetti,~\IEEEmembership{Member,~IEEE}
        and~Dario~Spiller
}

\markboth{PRE-PRINT. FINAL VERSION IN IEEE JOURNAL ON MINIATURIZATION FOR AIR AND SPACE SYSTEMS}
{A. Cratere \MakeLowercase{\textit{et al.}}: Efficient FPGA-accelerated Convolutional Neural
Networks for Cloud Detection on CubeSats}

\twocolumn[
\begin{@twocolumnfalse}
  
\maketitle

This is the pre-acceptance version. To read the final version published in the IEEE Journal on Miniaturization for Air and Space Systems, please go to: \highlightedurl{https://doi.org/10.1109/JMASS.2025.3533018} \\

\begin{abstract}
We present the implementation of four FPGA-accelerated convolutional neural network (CNN) models for onboard cloud detection in resource-constrained CubeSat missions, leveraging Xilinx’s Vitis AI (VAI) framework and Deep Learning Processing Unit (DPU), a programmable engine with pre-implemented, parameterizable IP cores optimized for deep neural networks, on a Zynq UltraScale+ MPSoC. This study explores both pixel-wise (Pixel-Net and Patch-Net) and image-wise (U-Net and Scene-Net) models to benchmark trade-offs in accuracy, latency, and model complexity. Applying channel pruning, we achieved substantial reductions in model parameters (up to 98.6\%) and floating-point operations (up to 90.7\%) with minimal accuracy loss. Furthermore, the VAI tool was used to quantize the models to 8-bit precision, ensuring optimized hardware performance with negligible impact on accuracy. All models retained high accuracy post-FPGA integration, with a cumulative maximum accuracy drop of only 0.6\% after quantization and pruning. The image-wise Scene-Net and U-Net models demonstrated strong real-time inference capabilities, achieving frame rates per second of 57.14 and 37.45, respectively, with power consumption of around 2.5 W, surpassing state-of-the-art onboard cloud detection solutions. Our approach underscores the potential of DPU-based hardware accelerators to expand the processing capabilities of small satellites, enabling efficient and flexible onboard CNN-based applications.

\end{abstract}


\end{@twocolumnfalse}]

{
  \renewcommand{\thefootnote}{}%
  \footnotetext{\textit{Corresponding author: Angela Cratere}.
  
Angela Cratere, Marcello Asciolla, Maria Rizzi and Francesco Dell’Olio, are with the Department of Electrical and Information Engineering, Polytechnic University of Bari, 70126 Bari, Italy (e-mails: a.cratere@phd.poliba.it, m.asciolla@phd.poliba.it, maria.rizzi@poliba.it, francesco.dellolio@poliba.it).

M. Salim Farissi, Augusto Nascetti, and Dario Spiller are with the School of Aerospace Engineering, Sapienza University of Rome, 00138 Rome, Italy (e-mails: mohamedsalim.farissi@uniroma1.it, augusto.nascetti@uniroma1.it, dario.spiller@uniroma1.it).

Andrea Carbone is with the Department of Astronautical, Electric and Energy Engineering, Sapienza University of Rome, 00184 Rome, Italy (e-mail: and.carbone@uniroma1.it).}}

%
\IEEEpeerreviewmaketitle

\section{Introduction}\label{sec:introduction}
\IEEEPARstart{T}he advent of \gls{ML}-based \gls{EC} solutions in space technology has introduced significant advancements in optimizing satellite data processing, enabling new opportunities to enhance onboard resource management and increasing in-orbit autonomy \cite{Murphy2021MachineLI, Ghiglione_2022, Ciardi_GPUSAT_2023}. Recently, \gls{DL} techniques have been proposed for several onboard applications, such as payload data processing, navigation and platform control \cite{Belokonov_2020-AI_attitude_CubeSat}, task scheduling \cite{aerospace10010078_task_scheduling_Zeleke} and \gls{FDIR}\cite{Horne_2023_anomaly_detection_NN_CubeSat}. Among these, the \gls{EO} applications hold particular promise, as \gls{AI} has the potential to enable real-time satellite imagery processing and to revolutionize its utilization. CubeSats offer a low-cost, rapid-development platform for experimenting with new \gls{AI}-driven approaches, which can increase the efficiency of \gls{EO} missions by reducing bandwidth utilization \cite{Guerrisi_2023} and enabling real-time alert services for time-sensitive applications such as emergency response -— a growing area of interest in the \gls{EO} community, as evidenced by the recent launch of \gls{ESA}'s $\Phi$-Sat-2 mission \cite{Phi_Sat_2}.

The growing use of \gls{ML} for resource-constrained space systems is further supported by advancements in \gls{HW} accelerators for small satellite and nanosatellite avionics \cite{George2018, Cratere_OBC_2024, Diana_2024_HW_accelerators}. Among these, \glspl{FPGA},
particularly when integrated into \gls{SoC} architectures, offer significant advantages in energy efficiency for data-intensive tasks compared to other \gls{COTS} \gls{HW} accelerators like \glspl{GPU} \cite{Qasaimeh_2021_accelerators_FPGA_energy_efficiency}. Moreover, \glspl{FPGA} provide a customizable architecture that can be optimized for specific processing needs, maximizing resource utilization. 
However, deploying \glspl{FPGA} as \gls{HW} accelerators presents unique challenges, requiring multiple development stages,  iterative model pruning, and low-level optimization, which often lead to extended development times \cite{Rapuano_2021_CloudScout_FPGA_vs_VPU}. Given the potentialities and intricacies of these \gls{HW} devices, this paper explores the implementation of \gls{CNN} models on \glspl{FPGA} for resource-constrained space systems, leveraging the \gls{DPU} microarchitecture within Xilinx's \gls{VAI} development environment \cite{Vitis_AI_v2.5_xilinx}. Specifically, we focused on EO on-board cloud detection as a case study and we tested four different \gls{CNN} models for \gls{MS} data analysis. This application is of considerable interest, as \gls{DL}-based cloud detection allows for the automatic in-orbit discarding of cloudy images, leading to substantial savings in onboard storage resources and bandwidth utilization for downlink \cite{Park_etal_2020}. The tested models include a one-dimensional (1D) and two-dimensional (2D) CNN for pixel-wise classification (Pixel-Net and Patch-Net, respectively), a U-Net for image segmentation, and an image-wise CNN for binary classification (Scene-Net). We deployed these models on an Avnet Ultra96-V2 board \cite{Ultra96V2}, equipped with a Xilinx Zynq UltraScale+ MPSoC, utilizing a \gls{DPU}-based \gls{HW} accelerator architecture. Their performance was compared to identify a benchmark \gls{CNN} architecture for \gls{DPU} deployment that balances accuracy, complexity, and latency, while carefully considering power consumption in the resource-constrained environment of CubeSats.

The two main contributions of this paper can be  summarized as follows:
\begin{itemize}
    \item We deployed multiple \gls{CNN} architectures on a \gls{DPU} engine, conducting a comparative analysis of different \gls{CNN} models for both segmentation and image-wise classification in the context of \gls{EO} onboard cloud detection application; 
    \item We provided insights into optimal \gls{CNN} architecture and deployment strategies which balance model complexity, accuracy, and resource consumption for real-time inference in resource-constrained environments.
\end{itemize}

The paper is organized as follows: \S \ref{section:methods} introduces the dataset used and describes the \gls{CNN} models and the \gls{FPGA} implementation process; \S \ref{sec:results} presents the experimental results; \S \ref{sec:discussion} discusses the major findings; and \S \ref{sec:conclusions} concludes the paper.

\begin{table*}[t!]
\centering
\caption{Overview of CNN Models for onboard cloud detection in the literature.}
\renewcommand{\arraystretch}{1.5}
\resizebox{1\textwidth}{!}{
\begin{tabularx}{\textwidth}{>{\raggedright\arraybackslash}X>{\centering\arraybackslash}X>{\centering\arraybackslash}X>{\centering\arraybackslash}X>{\centering\arraybackslash}X}
\toprule
\textbf{Study} & \textbf{\gls{NN} Architecture} & \textbf{Dataset} & \textbf{Application} & \textbf{HW -- Deployment Toolchain}\\ 
\toprule
Giuffrida \textit{et al.}, 2020 \cite{Giuffrida_2020_CloudScout_VPU} & CloudScout & 512$\times$512$\times$3 Sentinel-2 images  & Image-level binary cloud classification & VPU -- NCSDK\textsuperscript{*} \\ 
Rapuano \textit{et al.}, 2021 \cite{Rapuano_2021_CloudScout_FPGA_vs_VPU} & FPGA-CloudScout & - & Image-level binary cloud classification & Coarse-grained FPGA accelerator -- VHDL\\ 
Giuffrida \textit{et al.}, 2022 \cite{Giuffrida_2022_PhiSat1} & UNet-CloudScout & 192$\times$192$\times$3 synthetic HS data cubes (Sentinel-2) & Cloud segmentation & VPU -- NCSDK\textsuperscript{*} \\ 
Pitonak et al., 2022 \cite{Pitonak_2022_CloudSatNet1_FPGA_LandSat8} & CloudSatNet-1  & 512$\times$512$\times$3 Landsat-8 images & Image-level cloud classification&  Fine-grained FPGA accelerator -- FINN\\
Castelino \textit{et al.}, 2024 \cite{castelino2024energyefficientartefactdetectionaccelerator} & Unsupervised Convolutional AutoEncoder & Different 144$\times$144$\times$200 HS images& General Artefact Segmentation& DPU -- Vitis AI \\ 
\hline
This Study & Pixel-Net, Patch-Net, Scene-Net, U-Net & 256$\times$256$\times$12 Sentinel-2 images & Binary cloud classification and segmentation & DPU -- Vitis AI flow \\ 
\bottomrule
\multicolumn{5}{l}{\footnotesize \textsuperscript{*}Neural Compute Software Development Kit.}
\end{tabularx}}
\label{tab:comparison_previous_work}
\end{table*}

\subsection{Related Works}
Clouds are unavoidable in remote sensing imagery, particularly in optical bands, and can severely compromise the utility of satellite images for ground analysis, posing significant challenges for downlink operations, bandwidth optimization, and on-board resource utilization \cite{Park_etal_2020}. For this reason, on-board cloud detection is a critical theme in the EO field. The $\Phi$-Sat-1 was a pioneering effort in this direction, showcasing the first in-orbit deployment of a \gls{CNN} for cloud detection using a custom \gls{HW} accelerator based on the Movidius Myriad 2 \gls{VPU} \cite{Giuffrida_2022_PhiSat1}. In the original design, the \gls{CNN}, named \textit{CloudScout}, was a simple image-wise binary classifier used to distinguish between cloudy and not cloudy images \cite{Giuffrida_2020_CloudScout_VPU}. Later, the network was enhanced to resemble a U-Net architecture for segmentation tasks (\textit{UNet-CloudScout}\cite{Giuffrida_2022_PhiSat1}). The original CloudScout achieves notable results on a test dataset comprising 512$\times$512$\times$ 3 \gls{HS} datacubes synthesized from Sentinel-2 data, with an inference time of 325 ms, an accuracy of 92\%, a low false positive rate (1\%) and a total power consumption of 2 W per inference on the \gls{VPU}. 

After $\Phi$-Sat-1, some researchers investigated the possibilities of accelerating CNN cloud detection algorithms using other \gls{HW} accelerators, specifically FPGAs \cite{Rapuano_2021_CloudScout_FPGA_vs_VPU,Pitonak_2022_CloudSatNet1_FPGA_LandSat8}. For instance, \cite{Rapuano_2021_CloudScout_FPGA_vs_VPU} introduced an \gls{FPGA}-accelerated version of CloudScout (\textit{FPGA-CloudScout}) comparing the performance of the Myriad 2 VPU with a \gls{COTS} \gls{FPGA}-based \gls{HW} accelerator for the CloudScout case study. Their findings showed that while \glspl{FPGA} offer faster inference times (141.68 ms) and higher customization degree, they also result in higher power consumption (1.65 W for the accelerator alone and 3.4 W for the entire \gls{SoC} system) and longer development times.
Another notable effort is the \textit{CloudSatNet-1} introduced by \cite{Pitonak_2022_CloudSatNet1_FPGA_LandSat8}, an \gls{FPGA}-based quantized \gls{CNN} that, similarly to CloudScout, was designed for binary image-wise classification of cloudy/not cloudy images. Trained on the Landsat-8 Cloud Cover Assessment (CCA) validation RGB dataset, CloudSatNet-1 aimed to evaluate cloud classification performance across different biomes with varying cloud-to-terrain contrasts and to assess the impact of quantization on CNN accuracy. 
Findings indicate that using 512$\times$512 RGB images, accuracy reached 90\%, increasing to 94.4\% when excluding tiles with snow and ice, with a low false positive rate ($<$ 3\%). Quantization had minimal impact on the overall accuracy (around 2\% decrease for 3-bit and 4-bit models) but significantly reduced the memory footprint, enabling the deployment of the model on cost-effective \gls{FPGA} platforms. The tested Zynq-7020 \gls{SoC} achieves an average power consumption of 2.5 W.  
In addition to these \gls{FPGA}-based implementations, attempts to deploy U-Net architectures for cloud segmentation on CubeSats using traditional \glspl{muC} have been reported. For instance,  \cite{Park_2020_UNet_muP} used Sentinel-2 RGB images to train the \textit{Nano U-Net} (NU-Net). The network achieved 90\% accuracy with an inference time of 407.22 ms on the \gls{muC} for an RGB image of 48$\times$64 pixels. Similarly, \cite{Salzar_2022_UNet_muC} reported comparable accuracy (90\%) when evaluating NU-Net on images from the FACSAT-1 CubeSat.

While previous studies on \gls{FPGA}-based \gls{HW} accelerators have primarily focused on binary \glspl{CNN} for image-level cloud classification, this paper adopts a broader perspective by comparing various \gls{CNN} models for both classification and segmentation tasks. Additionally, this work explores \gls{DPU}-based architectures, which have not been previously investigated in the literature on in-orbit cloud detection. Notably, \cite{Rapuano_2021_CloudScout_FPGA_vs_VPU} used a coarse-grained \gls{FPGA} accelerator synthesized with VHDL, while \cite{Pitonak_2022_CloudSatNet1_FPGA_LandSat8} developed a fine-grained FPGA accelerator with FINN. To the best of our knowledge, the only other attempt to use the \gls{VAI}'s \gls{DPU} and development flow for onboard small satellite applications is the recent work by \cite{castelino2024energyefficientartefactdetectionaccelerator}, which investigates an unsupervised convolutional autoencoder (CAE) model for general artifact identification in HS images. Table \ref{tab:comparison_previous_work} provides an overview of related works and compares them with our approach.

\section{Material and methods}\label{section:methods}
We evaluated four different CNN models. The first, \textit{Pixel-Net}, is a one-dimensional (1D) model that makes inferences based exclusively on spectral features, without considering the spatial context of each pixel. This model was selected for its lightweight architecture and its effectiveness in processing \gls{MS} data \cite{Carbone_2024_GIS_conference}. The second model, \textit{Patch-Net}, integrates both spectral and spatial information within a 5$\times$5 neighborhood around each pixel, providing a more comprehensive analysis of the surrounding context. The third model is a customized version of a \gls{FCNN}, specifically a U-Net, which performs segmentation directly on images or tiles. To explore alternative approaches and enable a direct comparison of the DPU-based approach with previous FPGA implementations, a fourth model, \textit{Scene-Net}, was tested for image-level classification (cloudy/not cloudy). All models were trained using Sentinel-2 \gls{MS} data, as detailed in the next subsection.

\subsection{Dataset}\label{subsec:datset}

\begin{figure*}[htbp]
    \centering
    \begin{subfigure}[t]{0.495\textwidth}
        \centering
        \includegraphics[width=\textwidth]{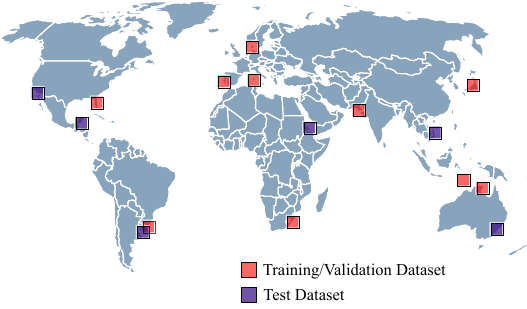}
        \caption{Global distribution of the datasets.}
        \label{subfig:subfigure_global_distribution_dataset}
    \end{subfigure}
    \hspace{0\textwidth} 
    \begin{subfigure}[t]{0.492\textwidth}
        \centering
        \includegraphics[width=\textwidth]{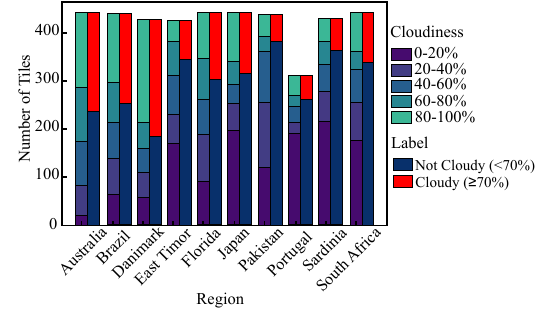}
        \caption{Cloudy percentages for training/validation dataset.}
        \label{fig:subfigure_cloudy_percentage_training_validation_dataset}
    \end{subfigure}
    \caption{(a) Global distribution of the Sentinel-2 imagery utilized for training, validation and test datasets. (b) Distribution of the cloudy percentage of the 256$\times$256 tiles obtained from Sentinel-2 granules used for the training and validation dataset. Tiles with cloudy percentage $\ge$ 70\% are labeled as cloudy in the construction of the dataset for the Scene-Net model.}
    \label{fig:distribution_cloudy_percentage_dataset}
\end{figure*}

The models were trained and evaluated using Sentinel-2 Level-2A (L2A) products \cite{Sentinel_SCL}, which provide \gls{BoA} reflectance values across 12 spectral bands, covering visible, near-infrared and short-wave infrared wavelengths. For this study, 16 scenes, captured between January 2022 and August 2024, were carefully selected from different global regions to provide a wide coverage of different Earth's surface types and ensure a balanced and representative dataset (Fig.~\ref{subfig:subfigure_global_distribution_dataset}). Of these scenes, 10 were chosen for the training and validation dataset, with a 70\% training and 30\% validation split, while the remaining 6 scenes constituted the test dataset. 
Since L2A products provide spectral data along various bands with different spatial resolutions (10 m, 20 m and 60 m), all bands were resampled to a common resolution of 20 m using bilinear interpolation. The \gls{SCL}, including 12 classes and provided at 20 m resolution within the L2A products, was used to obtain the cloud mask that served as ground truth for labeling the data. We constructed a labeled dataset by reorganizing the original 12 classes of the \gls{SCL} into \textit{cloudy} and \textit{not cloudy}, where the cloudy labels include the original high and medium probability cloud classes, and the not cloudy labels all other classes. Following a methodology similar to \cite{Carbone_2024_GIS_conference}, 20,000-pixel spectra were randomly sampled for each class, resulting in a dataset of 400,000 pixels for the Pixel-Net model. The Patch-Net dataset was similarly constructed by sampling 20,000 patches of 5$\times$5-pixels for each target label, with each patch labeled according to the class of its central pixel.  

The dataset for the Scene-Net and U-Net models was obtained by dividing the Sentinel-2 granules into tiles of 256$\times$256 pixels. Their size was determined by a trade-off between the model accuracy (and complexity) and the constraints of the used \gls{DPU} engine -- a DPUCZDX8G \gls{IP core} \cite{DPUCZDX8G_product_guide_xilinx} with a B1600 architecture \cite{PYNQ_DPU}, as detailed in \S\ref{subsec:fpga_deployment}. Initial tests using 512$\times$512 input images resulted in out-of-memory errors as the model size exceeded the \gls{DPU} capacity \cite{Vitis_AI_v2.5_xilinx}, leading to the adoption of the smaller tile size.  

To ensure data quality, tiles containing \textit{no data} pixels were excluded, as outliers can represent critical points during the training phase \cite{Giuffrida_2020_CloudScout_VPU}. For Scene-Net, each tile was labeled as \textit{cloudy} or \textit{not cloudy} based on a threshold of 70\% cloudy pixels (\cite{Giuffrida_2020_CloudScout_VPU,Rapuano_2021_CloudScout_FPGA_vs_VPU, Pitonak_2022_CloudSatNet1_FPGA_LandSat8}).

After preprocessing, the training and validation dataset comprised 4,231 tiles, with 2,976 classified as \textit{not cloudy} and 1,255 as \textit{cloudy}. The dataset was initially unbalanced toward the not cloudy class (Fig.~\ref{fig:distribution_cloudy_percentage_dataset}), which could have compromised the performance of the Scene-Net model, since an unbalanced dataset can increase model sensitivity to variations in input data distribution \cite{lecun2015deep_learning_review_nature}. To address this, we balanced the Scene-Net training and validation dataset by sampling an equal number of cloudy and not cloudy patches. To mitigate data scarcity and further enhance the robustness and generalization capability of the model, we employed an online data augmentation strategy, where augmentation was performed \textit{on-the-fly} \cite{malialis2022dataaugmentationontheflyactive} as the data was fed into the \gls{CNN}.

\subsection{Baseline Models Architectures}\label{sec:model_architectures}

\begin{figure*}[t]
    \centering
    \includegraphics[width=1\textwidth]{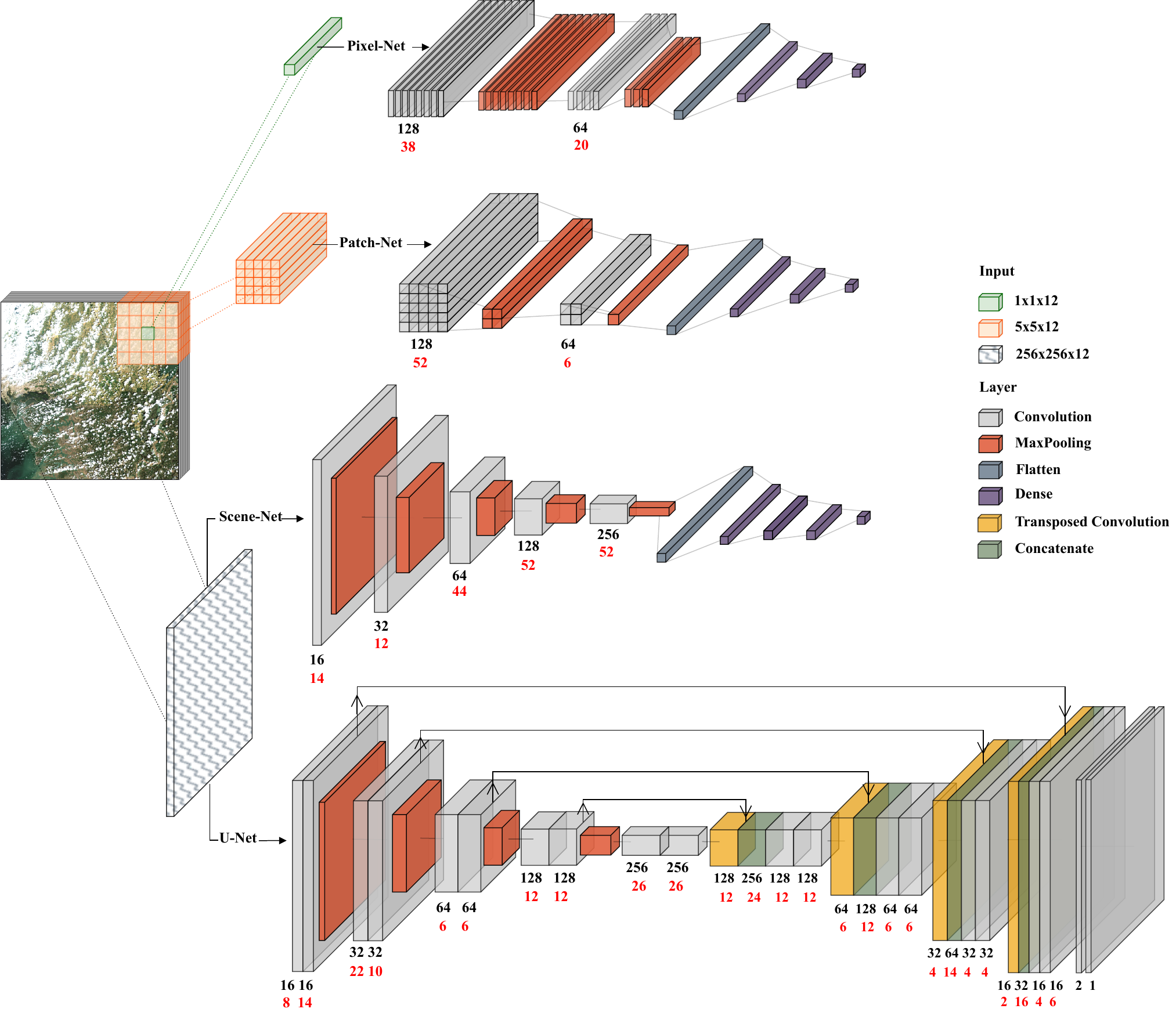}
    \caption{Architectures of the four CNN models (Pixel-Net, Patch-Net, Scene-Net, and U-Net) for cloud detection. Each model processes inputs of varying spatial resolutions and consists of multiple layers, including convolution, max-pooling, flatten, dense, transposed convolution, and concatenate layers, each represented by different colors. The black numbers indicate the number of convolutional kernels in the baseline models, while the red numbers show the number of filters after applying channel pruning.}
    \label{fig:models_architectures}
\end{figure*}

In this section, we detail the architecture of the four \gls{CNN} models, as shown in Fig.~\ref{fig:models_architectures}. Fine-tuning the final architectures required iterative modifications of the layers and hyperparameters. The objective was to maximize accuracy and reduce the number of \glspl{FP}, while minimizing the number of model parameters for optimal FPGA integration. One of the key requirements, following \cite{Giuffrida_2020_CloudScout_VPU}, was to maintain \glspl{FP} at the image level -- meaning non-cloudy images misclassified as cloudy 
according to the 70\% cloudy-pixel threshold -- around 1\%. This level is crucial because, in an operational mission where the models are used to decide which images to download to the ground and which can be discarded directly onboard the satellite, \glspl{FP} represent a net loss of useful data. The second essential requirement was \gls{HW} constraints, specifically, that the size of the activations associated with each layer cannot exceed the capacity of the DPU buffer memory, which is limited by a bank depth of 2048 units in the case of the B1600 architecture \cite{DPUCZDX8G_product_guide_xilinx}.

All CNNs were implemented in Python using the TensorFlow DL framework on a workstation equipped with an AMD Ryzen Threadripper PRO 7965WX 24-core CPU, 256 GB RAM, and an NVIDIA GeForce RTX 4090 GPU with 24 GB VRAM.

The models were trained with the Adam optimizer, using a learning rate of $10^{-3}$ and binary cross-entropy loss. Training was set to a maximum of 200 epochs, with early stopping applied based on a patience parameter of 30 epochs.

\subsubsection{Pixel-Net -- Pixel-wise 1D-CNN}\label{subsec:1D_CNN}

The Pixel-Net model was designed and implemented on \gls{FPGA} as an initial step, serving as a simplified \gls{CNN} architecture with low computational complexity and a minimal number of parameters, making it less complex to implement and deploy on \gls{HW}. The  architecture was inspired by \cite{Carbone_2024_GIS_conference}, which employed 1D operators commonly used in time-series and spectral data analysis. However, the original NN was adapted for \gls{FPGA} deployment by replacing traditional 1D convolutional and max-pooling layers with their 2D counterparts to ensure compatibility with the \gls{DPU}, which is primarily tailored for image processing tasks and does not directly support 1D operators \cite{DPUCZDX8G_product_guide_xilinx}. This modification enabled deployment on the \gls{HW} accelerator while maintaining the original functionality. Pixel-Net was further adapted for binary classification at the pixel level, using each pixel’s spectral signature as input (a 12-channel vector). The architecture consists of two convolutional layers with 128 and 64 filters, each with a kernel size of 3, same padding, ReLU activation, and followed by max pooling with a pool size of 2 and stride of 1. L2 regularization ($\lambda=10^{-5}$) is applied to prevent overfitting. After the convolutional layers, a flatten layer reshapes the output, which is then passed through two fully connected layers with 64 and 32 units, respectively, and a ReLU activation function. After the first dense layer, a dropout layer with a 20\% dropout rate is applied to further reduce overfitting. The output layer is a dense unit with a sigmoid activation function. \\

\subsubsection{Patch-Net -- Patch-wise 2D-CNN}
The Patch-Net model builds upon Pixel-Net by incorporating spatial information for each pixel. Its input consists of 5$\times$5-pixel patches with 12 spectral channels, using a sliding window approach across the image. The model architecture remains consistent with Pixel-Net for direct comparison, with the only modification being an increased dropout rate (50\%) to prevent overfitting.\\

\subsubsection{Scene-Net -- Image-wise CNN}\label{subsec:Imagebased_CNN} 
Scene-Net takes 256$\times$256-pixel tiles with 12 spectral channels as input and classifies entire tiles as cloudy or not cloudy using a deeper network architecture. The \gls{CNN} includes five convolutional layers with increasing filter sizes (16, 32, 64, 128, and 256 filters, respectively), each followed by ReLU activation and max-pooling layers that progressively reduce the spatial dimensions. The output is flattened and passed through three fully connected layers with 1024, 512, and 256 units, respectively, with dropout (50\%) applied between layers to reduce overfitting. The 1024-unit layer was added because it acts as a bottleneck, reducing activations and weight load for the DPU in the following layers, thus avoiding buffer overflow. The output layer consists of a dense unit with a sigmoid activation function. As stated in \S\ref{subsec:datset}, on-the-fly data augmentation techniques, including random rotations and horizontal and vertical flips of the images, were applied during training to improve generalization. \\

\subsubsection{U-Net -- Fully Convolutional Neural Network}\label{subsubsec:U_Net} 
This model performs segmentation to classify each pixel as cloudy or not cloudy using a U-Net architecture \cite{U_Net_original_paper_2015}, in which an encoder-decoder topology uses convolutional layers to progressively down-sample feature maps, followed by upsampling layers for reconstructing the spatial resolution. The input of the network consists of 256$\times$256-pixel tiles with 12 spectral channels. The layer structure, inspired by \cite{CloudGAN_2022}, differs from the original U-Net in that the number of filters in each convolutional layer is halved, reducing the number of model parameters to optimize \gls{HW} deployment. Furthermore, unlike \cite{CloudGAN_2022}, we used transposed convolution layers instead of standard upsampling, as this allows the model to learn more detailed features. The encoder comprises four convolutional blocks. Each block consists of two convolutional layers with filter sizes increasing from 16 to 128 filters, each followed by ReLU activation and max-pooling layers. L2 regularization ($\lambda=10^{-5}$) is applied to all layers to prevent overfitting, and a 50\% dropout rate is used in the last two blocks. At the bottleneck, two convolutional layers with 256 filters and ReLU activation are used, followed by dropout (50\%). The decoder mirrors the encoder structure, using transposed convolutional layers to upsample feature maps and skip connections to concatenate upsampled outputs with corresponding encoder features. Filter sizes decrease progressively from 128 to 16, and the final output layer is a 1$\times$1 convolution with sigmoid activation, producing a binary cloud mask.

\begin{figure}[t]
    \centering
    \includegraphics[width=0.5\textwidth]{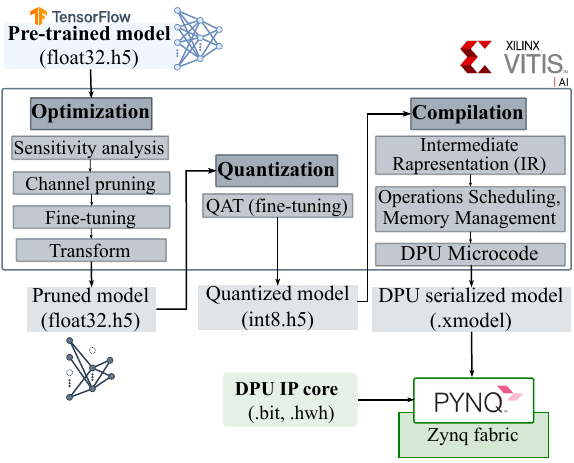}
    \caption{Overview of the CNN deployment strategy on FPGA using VAI and PYNQ frameworks.}
    \label{fig:Vitis_Ai_PYNQ_flowchart}
\end{figure}

\subsection{FPGA Deployment Strategy}\label{subsec:fpga_deployment}

The implementation of the \gls{CNN} models on the FPGA \gls{HW} accelerator used the \gls{VAI} tools, a comprehensive and powerful suite developed by Xilinx to facilitate \gls{AI} inference deployment on \gls{HW} platforms \cite{Vitis_AI_v2.5_xilinx}. The \gls{VAI} environment offers different components, including AI framework support, quantization and model compilation tools, and pre-compiled target \glspl{DPU}, i.e., specialized \gls{CNN} accelerators integrated as soft \glspl{IP core} for Xilinx \glspl{SoC}. The \gls{DPU} operates as a programmable engine purpose-built for \gls{CNN} inference, designed around a custom instruction set. 

Several key considerations guided the selection of the VAI toolchain and the DPU for model acceleration. Broadly, deep \gls{CNN} accelerators can be categorized into \textit{streaming} or \textit{application-specific} architectures and \textit{single computation engines} \cite{streaming_single_computation}. Streaming architectures, which include coarse-grained (e.g., System Verilog) and fine-grained (e.g., FINN or HLS4ML) designs, offer high optimization but require custom \gls{HW} configurations and synthesis tailored for each \gls{CNN} model, limiting flexibility. In contrast, the \gls{DPU} exemplifies a single computation engine that processes \gls{CNN} layers sequentially. This approach provides a more adaptable, less complex—and therefore less error-prone—solution, since the \gls{DPU}, once configured, can support multiple \gls{CNN} architectures without requiring recompilation or programmable logic reconfiguration. This adaptability is crucial, especially for CubeSat missions \cite{Cratere_OBC_2024}, as it allows onboard computing units to scale according to dynamic mission requirements and adapt to changing tasks without additional \gls{HW} customization, thereby reducing time-to-market and complexity. Additionally, \cite{Machura_2022_FINN_vs_VITIS_AI} demonstrated that \gls{DPU}-based solutions offer higher accuracy, lower power consumption, and reduced \gls{FPGA} resource usage, particularly memory, compared to application-specific designs for similar \gls{CNN} configurations.

For this study, the DPUCZDX8G IP core was chosen for its compatibility with Zynq UltraScale+ MPSoC devices \cite{DPUCZDX8G_product_guide_xilinx}, specifically, the B1600 architecture -- nomenclature denoting the number of \gls{MAC} operations per clock cycle.

The FPGA deployment using the \gls{VAI} development toolchain consisted of four primary stages, summarized in Fig.~\ref{fig:Vitis_Ai_PYNQ_flowchart}.

\begin{itemize}
\item \textit{Optimization}. This initial stage introduces \textit{sparsity} \cite{hoefler2021sparsity} through structured pruning \cite{Li2016PruningFF} to minimize the model’s computational load and memory requirements by reducing the model’s parameters and \gls{FLOPs}. Specifically, we employed the \gls{VAI} coarse-grained pruning algorithm to achieve channel (or convolutional filter) sparsity \cite{Li2016PruningFF} within each layer, following four main steps: first, sensitivity analysis identifies and prioritizes low-impact channels for pruning based on a target \gls{PR}, which specifies the target reduction in \gls{FLOPs} and parameters; next, iterative pruning is applied to remove the selected channels to gradually reach the desired sparsity, followed by fine-tuning to recover any loss in accuracy. Finally, the model is transformed from a sparse to a dense form with inactive channels permanently removed, resulting in a pruned model ready for quantization. 

\item \textit{Quantization}. This is an essential stage to reduce the memory footprint of \gls{DL} models and to allow them to be integrated on \glspl{DPU}. Model training typically yields \gls{float32} weights and activations, whereas the Xilinx \glspl{DPU} support only \gls{int8} arithmetic. We adopted a \gls{QAT} strategy \cite{krishnamoorthi2018quantizingdeepconvolutionalnetworks}, an approach that introduces low-precision constraints during training, allowing the model to learn and adapt to \gls{int8} precision and to minimize the accuracy loss that might occur with \gls{PTQ}. During \gls{QAT}, the model’s forward pass simulates \gls{int8} constraints on weights and activations while maintaining \gls{float32} precision in gradients during backpropagation. Our implementation used the \gls{VAI} quantizer with a per-tensor quantization strategy.

\item \textit{Compilation}. After \gls{QAT}, the models were compiled using the \gls{VAI} compiler, which translates the quantized model (in \texttt{.h5} format) into \gls{DPU}-executable instructions (\texttt{.xmodel} format). The compiler first converts the input model into an \gls{IR} format that organizes the model as a graph of operations optimized for \gls{HW} deployment, applying instruction scheduling to leverage \gls{DPU} parallelism. The \gls{IR} is then translated into a \gls{DPU}-specific microcode and serialized as a \texttt{.xmodel} file, which contains the quantized weights, optimized instructions, and memory allocation for efficiently handling \gls{HW} inference.

\item \textit{Model deployment and execution}. The models were deployed using PYNQ \cite{PYNQ_zynq}, which provides Python-based APIs and libraries to interact with \gls{FPGA} \gls{HW} and facilitate the deployment of \gls{DL} models in the \gls{DPU} core \cite{PYNQ_DPU}. PYNQ simplifies the seamless execution of \gls{DL} models by managing the data flow between the \gls{CPU} and the \gls{DPU} in the Zynq fabric. During inference, the \gls{DPU} handles the bulk of inference computations while the \gls{CPU} manages data input/output and pre- and post-processing tasks, allowing to achive a balance between inference speed and power efficiency.

\end{itemize}

The model optimization and pruning process was a very fundamental step in this development flow, allowing us to reduce model parameters significantly and improve resource usage efficiency. In our implementation, we experimented with \glspl{PR} from 0.1 to 0.9 across models to find the optimal trade-off between parameter reduction, accuracy retention, and control over \glspl{FP}. For all models except Scene-Net, a \gls{PR} of 0.9 achieved minimal accuracy loss, as detailed in \S \ref{sec:results}. For Scene-Net, the optimal \gls{PR} was selected to keep \glspl{FP} at 1\%. Since values of \glspl{PR} $>$ 0.6 resulted in \glspl{FP} of 2-3\%, the ideal \gls{PR} for this model was set at 0.6. This pruning process reduced the convolutional filters across layers by up to 90\% in some cases, as shown in Fig.~\ref{fig:models_architectures}. The number of parameters and \gls{FLOPs} for each model, both pre- and post-pruning, are detailed in Tab.~\ref{table:parameters_flops}.

\begin{table}[t!]
\renewcommand{\arraystretch}{1.2}
\centering
\caption{Number of parameters and \gls{FLOPs} before and after model pruning.}
\resizebox{0.5\textwidth}{!}{ 
\begin{tabular}{llcc}
\toprule
\textbf{Model} & & \textbf{Number of Parameters} & \textbf{FLOPs} \\
\toprule
\textbf{Pixel-Net} & \textbf{Baseline} & 68.29K & 642.40K \\
          & \textbf{Pruned}  & 17.43K & 84.67K \\

\textbf{Patch-Net} & \textbf{Baseline} & 94.02K & 1.30M \\
          & \textbf{Pruned}   & 13.00K & 380.94K \\

\textbf{U-Net}     & \textbf{Baseline} & 1.94M & 6.28G \\
          & \textbf{Pruned}   & 26.62K & 579.60M \\

\textbf{Scene-Net} & \textbf{Baseline} & 13.90M & 806.32M \\
          & \textbf{Pruned}  & 3.32M & 338.76M \\
\bottomrule
\end{tabular}
}
\label{table:parameters_flops}
\end{table}

\section{Results}\label{sec:results}
\subsection{Segmentation and Classification Performance}

\begin{table*}[t!]
\renewcommand{\arraystretch}{1.2}
\centering
\caption{Comparison of baseline (\gls{float32}) and pruned quantized \gls{FPGA}-deployed (\gls{int8}) models: accuracy, parameter and \gls{FLOPs} reduction.}
\resizebox{1\textwidth}{!}{ 
\begin{tabular}{lcccc}
\toprule
\textbf{Model} & \textbf{\gls{float32} Baseline Accuracy (\%)} & \textbf{\gls{int8} Pruned Accuracy (\%)} & \textbf{Parameters Reduction (\%)} & \textbf{\gls{FLOPs} Reduction (\%)} \\
\toprule
\textbf{Pixel-Net} & 95.94& \textbf{95.71} &   74.5 &  88.8\\ 

\textbf{Patch-Net} & 97.55 & \textbf{97.42} &   86.2 &   70.7\\ 

\textbf{U-Net}     & 98.46 & \textbf{98.04}&   98.6 &   90.7\\ 

\textbf{Scene-Net} & 99.0& \textbf{98.4}&   76.1 &   57.9\\ 
\bottomrule
\end{tabular}
}
\label{tab:evaluation_metrics_float32_int8_comparison}
\end{table*}

\begin{table*}[h]
\renewcommand{\arraystretch}{1.2}
    \centering
    \caption{Confusion matrices of the FPGA-deployed CNN models on the test datasets.}\label{tab:confusion_matrix}

    \begin{subtable}{0.24\textwidth}
        \centering
        \caption{\textbf{Pixel-Net}}\label{subtab:confusion_matrix_PixelNet}
        \begin{tabular}{cc|c} 
            \cline{2-3}
            Not Cloud & 97.79\% & 2.21\% (FP) \\
            \cline{2-3}
            Cloud  & 6.37\% (FN) & 93.63\% \\
            \cline{2-3}
             & \multicolumn{1}{c}{\rotatebox{45}{Not Cloud }} &\multicolumn{1}{c}{\rotatebox{45}{Cloud}}
        \end{tabular}
    \end{subtable}%
    \hspace{0.05\textwidth} 
    \begin{subtable}{0.23\textwidth}
        \centering
        \caption{\textbf{Patch-Net}}\label{subtab:confusion_matrix_PatchNet}
        \begin{tabular}{cc|c} 
            \cline{2-3}
             & 97.30\% & 2.70\% (FP) \\
            \cline{2-3}
             & 2.48\% (FN) & 97.52\% \\
            \cline{2-3}
             & \multicolumn{1}{c}{\rotatebox{45}{Not Cloud }} & \multicolumn{1}{c}{\rotatebox{45}{Cloud}}
        \end{tabular}
    \end{subtable}%
    \hspace{0.00000001\textwidth} 
    \begin{subtable}{0.23\textwidth}
        \centering
        \caption{\textbf{U-Net}}\label{subtab:confusion_matrix_UNet}
        \begin{tabular}{cc|c} 
            \cline{2-3}
             & 99.16\% & 0.84\% (FP) \\
            \cline{2-3}
             & 2.83\% (FN) & 97.17\% \\
            \cline{2-3}
             & \multicolumn{1}{c}{\rotatebox{45}{Not Cloud }} & \multicolumn{1}{c}{\rotatebox{45}{Cloud}}
        \end{tabular}
    \end{subtable}%
    \hspace{0.00000001\textwidth} 
    \begin{subtable}{0.23\textwidth}
        \centering
        \caption{\textbf{Scene-Net}}\label{subtab:confusion_matrix_SceneNet}
        \begin{tabular}{cc|c} 
            \cline{2-3}
            & 99.10\% & 0.90\% (FP) \\
            \cline{2-3}
            & 2.15\% (FN) & 97.85\% \\
            \cline{2-3} 
             & \multicolumn{1}{c}{\rotatebox{45}{Not Cloud }} & \multicolumn{1}{c}{\rotatebox{45}{Cloud}}
        \end{tabular}
    \end{subtable}%

\end{table*}

\begin{table*}[ht!]
\centering
\caption{CNN models characterization on FPGA board.}
\renewcommand{\arraystretch}{1.5}
\resizebox{\textwidth}{!}{
\begin{tabularx}{\textwidth}{>{\raggedright\arraybackslash}X>{\centering\arraybackslash}X>
{\centering\arraybackslash}X>
{\centering\arraybackslash}X>
{\centering\arraybackslash}X>{\centering\arraybackslash}X>{\centering\arraybackslash}X>{\centering\arraybackslash}X>{\centering\arraybackslash}X}
\toprule 
& \multicolumn{2}{c}{\textbf{Pixel-Net}} & \multicolumn{2}{c}{\textbf{Patch-Net}} & \multicolumn{2}{c}{\textbf{U-Net}}&\multicolumn{2}{c}{\textbf{Scene-Net}} \\ 
\cmidrule(lr){2-3} \cmidrule(lr){4-5} \cmidrule(lr){6-7} \cmidrule(lr){8-9}
 & \textbf{Baseline} &  \textbf{Pruned} & \textbf{Baseline}& \textbf{Pruned}  & \textbf{Baseline} & \textbf{Pruned} & \textbf{Baseline} & \textbf{Pruned} \\ 
\toprule
\textbf{Inference Time (ms)} & 0.35/px& 0.30/px & 0.36/px &0.31/px & 55.9 & 26.7 & 24.3  & 17.5\\ 

\textbf{Frame-per-second} & 0.045& 0.051   & 0.042 & 0.049 & 17.89& 37.45 & 41.15& 57.14\\ 

\textbf{Power Consumption (W)} &2.4 & 2.4& 2.6 & 2.5 & 2.7 & 2.4 & 3& 2.5 \\ 
\bottomrule 
\end{tabularx}}
\label{tab:cnn_performance_FPGA}
\end{table*}

We assessed the two pixel-wise models (Pixel-Net and Patch-Net) on a test dataset of 240,000 pixels and the two image-wise models (Scene-Net and U-Net) on a subset of the initial test dataset consisting of 500 tiles, due to the memory limitations of the Ultra96-V2 board (2GB RAM). To assess the effects of quantization, pruning, and \gls{FPGA} deployment, we analyzed each model's performance in both its baseline (\gls{float32} precision) and \gls{HW}-deployed (\gls{int8} representation) forms, as well as before and after pruning.

After \gls{FPGA} integration, all models retained strong performance metrics in their pruned versions, with Pixel-Net and Patch-Net achieving 95.7\% and 97.4\% accuracy, respectively, and U-Net and Scene-Net reaching 98\% and 98.4\% on their respective test datasets. \gls{QAT} effectively limited any accuracy drop, with reductions of about 0.1-0.3\%  across the models. This minor impact aligns with prior findings in the literature -- for instance, \cite{Rapuano_2021_CloudScout_FPGA_vs_VPU} reported a 0.3\% drop in accuracy after quantizing from \gls{float32} to \gls{int8}, while \cite{Pitonak_2022_CloudSatNet1_FPGA_LandSat8} observed a higher decrease of 3-6\% when converting from \gls{float32} to 4- and 2-bit integer models -- suggesting that \gls{QAT} effectively mitigates accuracy degradation due to the \gls{int8} representation.

Pruning introduced a further small accuracy drop of about 0.1-0.3\%, depending on the model, while significantly lowering the number of parameters and \gls{FLOPs} (Table \ref{table:parameters_flops}), with reductions of up to 98.6\% in parameters and 90.7\% in \gls{FLOPs} in the case of the U-Net model. The combined effects of quantization and structured pruning thus preserved high performance, underscoring the effectiveness of these techniques for achieving efficient \gls{FPGA} deployment without compromising accuracy significantly. The accuracy of the baseline (\gls{float32}) and final pruned (\gls{int8}) models, along with the reductions in parameters and \gls{FLOPs}, are summarized in Table \ref{tab:evaluation_metrics_float32_int8_comparison}.

The normalized confusion matrices in Tables \ref{subtab:confusion_matrix_PixelNet}--\ref{subtab:confusion_matrix_UNet} provide an in-depth look at each \gls{CNN} model's performance on the test dataset in their pruned, \gls{FPGA}-deployed versions. Both pixel-wise models (Pixel-Net and Patch-Net) effectively distinguish cloudy from non-cloudy pixels, with \glspl{FP} of around 2-3\%. Pixel-Net misclassified approximately 6\% of cloudy pixels as non-cloudy, while Patch-Net reduced \glspl{FN} to about 2.5\%. The U-Net model further improves these metrics, achieving \glspl{FP} close to 1\% and \glspl{FN} around 3\%.

\begin{figure*}[htbp]
    \centering
    \includegraphics[width=1\textwidth]{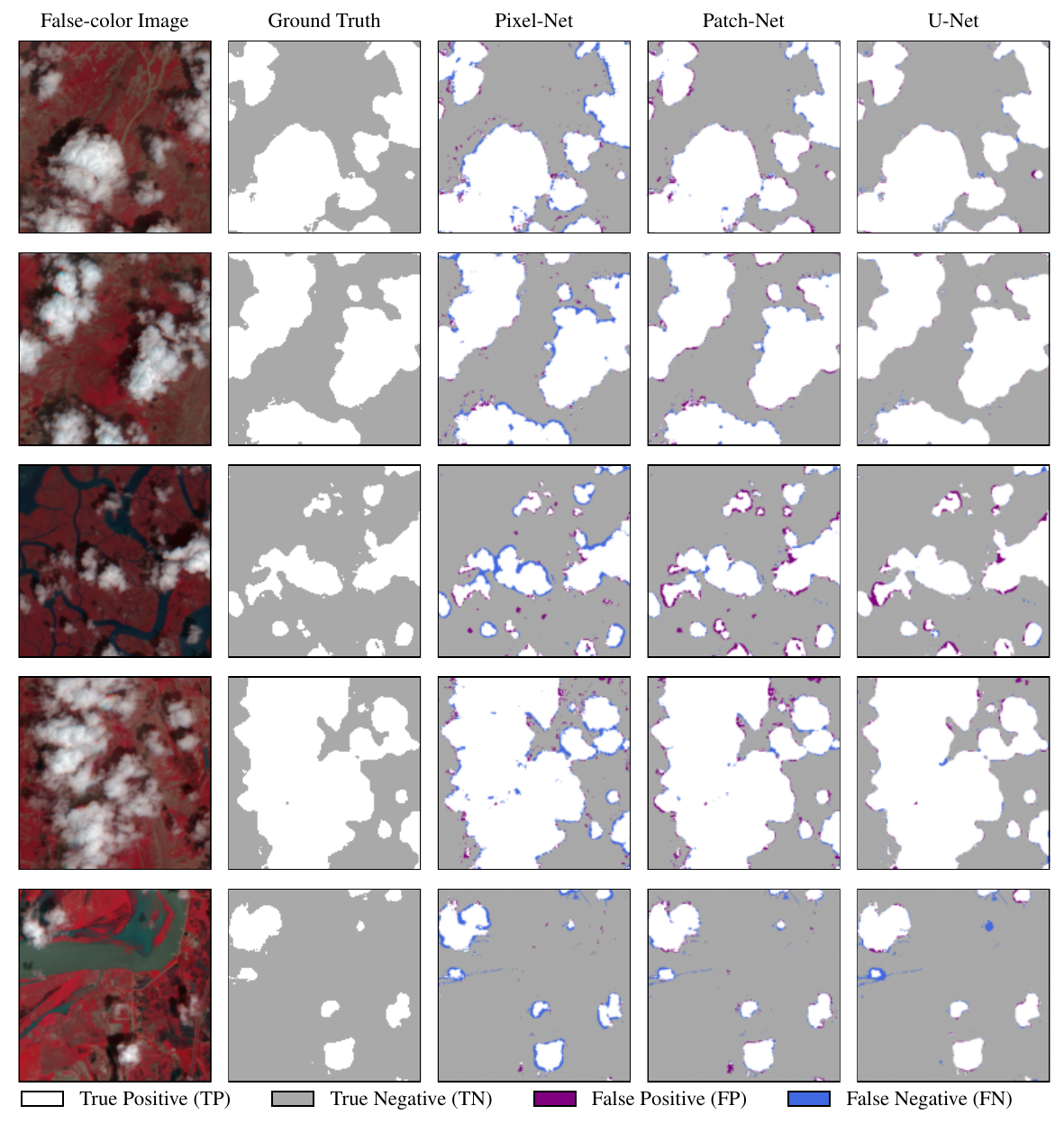}
    \caption{FPGA segmentation outputs for five different regions (rows). The five columns show: (1) the false-color RGB image (using B11, B3, B2 bands); (2) the ground truth cloud mask derived from the Sentinel-2 Scene Classification Layer (SCL); (3), (4), (5) the FPGA prediction from Pixel-Net, Patch-Net, and U-Net, respectively. }
    \label{fig:segmentation_results}
\end{figure*}

A visual examination of \gls{FPGA} segmentation outputs (Fig.~\ref{fig:segmentation_results}) supports these findings, suggesting that the 2D models (Patch-Net and U-Net), which incorporate spatial context, capture cloud boundaries more accurately than the 1D model (Pixel-Net), which tends to misclassify cloud pixels in transitional regions where clouds blend with land or shadows. It is worth noting that, for these three models, the metrics in Tables \ref{tab:evaluation_metrics_float32_int8_comparison} and \ref{subtab:confusion_matrix_PixelNet}--\ref{subtab:confusion_matrix_UNet} reflect segmentation algorithm performance. For Pixel-Net and Patch-Net, pixel-level \glspl{FP} of 2\% and 3\% correspond to tile-level \gls{FP} rates of 1.36\% and 0.9\%, respectively. U-Net’s segmentation accuracy of 98.04\% yields a tile-level accuracy of 98.4\%, with a very low \gls{FP} rate of 0.1\%. Thus, all three models meet the requirement of maintaining tile-level \glspl{FP} no more than 1\%.

For Scene-Net, results in Table \ref{subtab:confusion_matrix_SceneNet} confirm its strong classification capability, correctly identifying 99\% of cloudy tiles with an \gls{FP} rate around 1\% and minimal misclassification between cloudy and non-cloudy tiles.

\subsection{FPGA Deployment}

Table \ref{tab:cnn_performance_FPGA} provides a characterization of the four models, both in their baseline and pruned versions, on the \gls{FPGA} \gls{HW} accelerator in terms of inference time, \gls{FPS} and power consumption. Inference times were measured using an internal counter triggered by a start signal from the CPU and stopped at the end of inference. Real-time monitoring of power consumption was conducted via the \texttt{pmbus} interface on the Ultra96-V2 board, with the PYNQ framework providing access to the power rails \cite{pynq_pmbus}.

Baseline Pixel-Net and Patch-Net exhibited fast pixel-wise classification with average inference times of 0.35 ms/px and 0.36 ms/px, respectively. However, processing a complete 256$\times$256 tile required over 20 s for both models, corresponding to frame rates of 0.045 \gls{FPS} and 0.042 \gls{FPS}. After pruning, both inference times improved by 14\%, though still insufficient for efficient full-image processing.
In contrast, the Scene-Net and U-Net models demonstrated significantly faster inference times of 24.3 ms and 55.9 ms per 256$\times$256 tile, corresponding to frame rates of approximately 41.15 \gls{FPS} and 17.89 \gls{FPS}. Post-pruning, Scene-Net’s inference time decreased by 28\% (from 24.3 ms to 17.5 ms), and U-Net saw a 52\% reduction (from 55.9 ms to 26.7 ms), raising their frame rates to 57.14 \gls{FPS} and 37.45 \gls{FPS}, respectively, showcasing real-time classification capabilities.
Before pruning, all models exhibited average power consumption around 2.4-3 W, approximately 0.25-0.75 W above the DPU’s idle state. Pruning reduced power consumption by up to 17\%, notably in the case of the U-Net model.

\section{Discussion}\label{sec:discussion}

The performance comparison of our four \gls{DPU}-accelerated \gls{CNN} models reveals unique advantages and limitations associated with each model architecture, particularly when considering their suitability in the cloud detection scenario. For all models, the \gls{DPU} power consumption values (Tab.~\ref{tab:cnn_performance_FPGA}) align well with the typical power budgets of CubeSats -- 1-3 W for 1U, 2-5 W for 2U, and 7-20 W for 3U configurations \cite{Arnold2012EnergyBF} -- making \gls{DPU}-based solution highly suitable for nanosatellite applications. 

Pixel-wise models (Pixel-Net and Patch-Net) face latency challenges when applied to full-image segmentation. In our tests, segmenting a 256$\times$256 image with these models took approximately 20 s —- an impractically long time for real-time in-orbit cloud detection (for comparison, UNet-CloudScout on $\Phi$-Sat-1 required 181 ms \cite{Giuffrida_2022_PhiSat1}). 
Extending this to an entire 1152$\times$1152 HyperScout image would require around 450 s, whereas the UNet-CloudScout would require just 4 s to process similar number of pixels (see Tab.2). Despite the significant reduction \gls{FLOPs} achieved through pruning, the pixel-wise models did not gain sufficient latency improvements due to the \gls{DPU}’s lack of batch processing optimization, which forces throughput to be the inverse of latency. In contrast, the Scene-Net and U-Net models, which directly operate on entire tiles, showed significantly faster performance (\gls{FPS} of 37.45 and 57.14, respectively), while maintaining power consumption levels around 2.5 W -- making these models the most promising option for operational deployment in space missions. 

As shown in Table \ref{tab:comparison_with_literature}, which summarizes the comparison of our results with previous works in the literature, notably our \gls{DPU}-deployed pruned Scene-Net model demonstrates substantial improvements over the \gls{VPU}-accelerated CloudScout classifier in terms of accuracy (98.4\% for our model vs. 92\% for CloudScout) and latency (57.14 \gls{FPS} for our model vs. 12.31 \gls{FPS} for CloudScout), albeit with slightly higher power consumption (2.5 W vs. CloudScout’s 1.8 W). Our implementation also shows favorable performance compared to other \gls{FPGA}-accelerated cloud detection models: specifically, Scene-Net exhibited comparable accuracy and power consumption to CloudSatNet-1 (accelerated on a fine-grained FINN accelerator) while providing significant improvements over the FPGA-CloudScout model (VHDL  coarse-grained accelerator), achieving both lower power consumption and higher throughput (57.14 \gls{FPS} for our model vs. 28.23 \gls{FPS} for FPGA-CloudScout). Similarly, our U-Net model outperforms the UNet-CloudScout network, achieving a significantly higher throughput (37.45 \gls{FPS} for our U-Net vs. 5.51 \gls{FPS} for UNet-CloudScout) while maintaining similar accuracy. 

These results illustrate that our \gls{DPU}-based implementation strikes an optimal balance among accuracy, power efficiency, and resource utilization, achieving faster inference times and a more streamlined deployment process compared to both coarse-grained and fine-grained \gls{FPGA} solutions in the literature. The reduced resource demand aligns seamlessly with CubeSat constraints and highlights the scalability of similar architectures for other satellite missions that could benefit from onboard cloud detection or related segmentation tasks, where \gls{DL} processing onboard can significantly enhance mission capabilities.

\begin{table*}[t!]
\renewcommand{\arraystretch}{1.2}
\centering
\caption{Comparative performance of \gls{HW}-accelerated cloud detection \gls{CNN} in literature versus our models.}
\resizebox{1\textwidth}{!}{ 
\begin{tabular}{lcccc}
\toprule
\textbf{Model -- HW} & \textbf{Accuracy (\%)} & \textbf{Inference Time (ms)} & \textbf{Frame-per-second} & \textbf{Power Consumption (W)} \\
\toprule
CloudScout -- VPU \cite{Giuffrida_2020_CloudScout_VPU}& 92.0& 81.3* &   12.3 &  1.8\\ 

FPGA-CloudScout -- FPGA \cite{Rapuano_2021_CloudScout_FPGA_vs_VPU} & -- & 35.4* &   28.2 &   3.4\\ 

CloudSatNet-1 -- FPGA \cite{Pitonak_2022_CloudSatNet1_FPGA_LandSat8}    & 94.84 & --&   -- &   2.5\\ 

\textbf{Our pruned Scene-Net -- DPU} &\textbf{98.40}&   \textbf{17.5} &   \textbf{57.1} & \textbf{2.5}\\ 
\hline
UNet-CloudScout -- VPU \cite{Giuffrida_2022_PhiSat1}    & 95.10 & 181.3*&   5.5 &   1.8\\ 

\textbf{Our pruned U-Net -- DPU} & \textbf{99.0}& \textbf{26.7}&   \textbf{37.5}&   \textbf{2.4}\\ 
\bottomrule
*Values normalized to 256$\times$256 pixels for easier comparison.
\end{tabular}
}
\label{tab:comparison_with_literature}
\end{table*}

\section{Conclusions}\label{sec:conclusions}

This paper presents the \gls{DPU} implementation of four distinct \gls{CNN} models for CubeSat applications, focusing specifically on onboard cloud detection as a case study. The primary objective of this study was to assess the suitability of these models when embedded in \gls{DPU}-based \gls{HW} accelerators, evaluating the potential of the \gls{DPU} architecture to address the unique challenges of \gls{AI} applications in CubeSat missions. This study specifically investigated the trade-offs between pixel-wise and image-wise models in terms of accuracy, latency, and resource efficiency. All the tested models achieved high performance on the test dataset, with minimal accuracy drop after \gls{HW} deployment. Pixel-wise models (Pixel-Net and Patch-Net), while achieving high segmentation accuracy, faced latency limitations when applied to full-image analysis due to extended inference times. The pruning strategy applied to these models led to substantial reductions in \gls{FLOPs}, although it did not significantly reduce latency. Image-wise models (Scene-Net and U-Net) offered substantial speed (\gls{FPS} of 57.14  and  37.45, respectively) while maintaining low power consumptions (2.5 W), outperforming existing onboard cloud detection implementations in the literature. 

This study also emphasizes the inherent trade-offs associated with image-wise models, where model complexity must be carefully balanced against \gls{HW} capabilities to not exceed \gls{DPU} constraints. On the other hand, pixel-based models provide segmentation tasks with relatively more lightweight architectures, though their applicability is limited to scenarios where in-orbit inference could be limited to specific pixels, as in event-based detection -- e.g., flood \cite{Gracia_2021_flood_mapping_smallsats} and wildfire \cite{rs15030720} mapping, where the use of dedicated spectral indices can focus inference on specific pre-selected pixels. 
We wish to emphasize that image-level models generally require a substantial dataset for training; pixel-based models are the only viable solution when the dataset for training is limited (e.g. \cite{rs15030720}); thus, investigating their performance on \gls{FPGA} was nevertheless of interest to assess their practical application. 

To conclude, the deployment of \gls{CNN} models in a \gls{DPU} engine demonstrated the viability of this design choice in effectively accelerating \glspl{CNN} on \glspl{FPGA} while maintaining low power consumption. Future work will investigate the use of homogeneous and heterogeneous multi-core \glspl{DPU} \cite{10247793_Heterogeneous_multi_DPU_engines} to further enhance processing capabilities.

\bibliographystyle{IEEEtran}
\bibliography{bibliography}

\end{document}